\documentclass[journal]{IEEEtran}
\usepackage{cite}

\ifCLASSINFOpdf
\else
\fi
\usepackage{amsmath}
\usepackage{physics}
\usepackage{graphicx}
\usepackage{gensymb}

\usepackage[group-separator={,}]{siunitx}
\usepackage{lineno}

\DeclareRobustCommand*{\rmark}[1]{\raisebox{0pt}[0pt][0pt]{\textsuperscript{\tiny\ensuremath{\ifcase#1\or *\or \dagger\or \ddagger\or%
    \mathsection\or \mathparagraph\or \|\or **\or \dagger\dagger%
    \or \ddagger\ddagger \else\textsuperscript{\expandafter\romannumeral#1}\fi}}}}

\begin{document}
%
\title{Scaling up Superconducting Quantum Computers\\ with Cryogenic RF-photonics}
%
%
%

\author{Sanskriti Joshi,
        Sajjad Moazeni,~\IEEEmembership{Member,~IEEE}
        
\thanks{S. Joshi and S. Moazeni are with the Department
of Electrical and Computer Engineering, University of Washington, Seattle,
WA, 98195 USA e-mails: sjoshi3@uw.edu, smoazeni@uw.edu.}

\thanks{Manuscript received April 19, 2005; revised August 26, 2015.}}

%
%

\markboth{Journal of \LaTeX\ Class Files,~Vol.~14, No.~8, August~2015}%
{Shell \MakeLowercase{\textit{et al.}}: Bare Demo of IEEEtran.cls for IEEE Journals}
%



\maketitle

\begin{abstract}

Today's hundred-qubit quantum computers require a dramatic scale up to millions of qubits to become practical for solving real-world problems. Although a variety of qubit technologies have been demonstrated, scalability remains a major hurdle. Superconducting (SC) qubits are one of the most mature and promising technologies to overcome this challenge. However, these qubits reside in a millikelvin cryogenic dilution fridge, isolating them from thermal and electrical noise. They are controlled by a rack-full of external electronics through extremely complex wiring and cables. Although thousands of qubits can be fabricated on a single chip and cooled down to millikelvin temperatures, scaling up the control and readout electronics remains an elusive goal. This is mainly due to the limited available cooling power in cryogenic systems constraining the wiring capacity and cabling heat load management.

In this paper, we focus on scaling up the number of XY-control lines by using cryogenic RF-photonic links. This is one of the major roadblocks to build a thousand qubit superconducting QC. We will first review and study the challenges of state-of-the-art proposed approaches, including cryogenic CMOS and deep-cryogenic photonic methods, to scale up the control interface for SC qubit systems. We will discuss their limitations due to the active power dissipation and passive heat leakage in detail. By analytically modeling the noise sources and thermal budget limits, we will show that our solution can achieve a scale up to a thousand of qubits. Our proposed method can be seamlessly implemented using advanced silicon photonic processes, and the number of required optical fibers can be further reduced by using wavelength division multiplexing (WDM).

\end{abstract}


%
\IEEEpeerreviewmaketitle

\section{Introduction}
\label{sec:introduction}

\IEEEPARstart{T}{he} realization of quantum computers (QC) will revolutionize fields from physics and medicine to artificial intelligence and cryptography by exponentially speeding up certain computational classes~\cite{Biamonte2017, Egger2020, Preskill2018}. While quantum supremacy has recently been claimed using 53 superconducting qubits~\cite{Arute2019}, practical quantum computers for real-world applications require a dramatic scale-up to millions of qubits~\cite{Brooks2019}. Among various qubit technologies, superconducting (SC) qubits are still the most mature and promising technology for realizing such large-scale QCs due to the many advantages they offer: ease of fabrication using solid-state processes, fast gate operation with high fidelity, and use of SC circuits and interconnects~\cite{Hofheinz2009, Foxen2018, McDermott2018, Bronn2018}. Today, SC qubits enabled the world's largest QC with 128 qubits~\cite{rigetti-128qbit,ibm-128qbit}. These SC qubits must be isolated from thermal and electrical noise sources by placing them at millikelvin temperatures (deep cryogenic) in a dilution refrigerator (an example shown in Fig.~\ref{fig:transmonqubit}a). This makes communications with the qubits extremely challenging. External electronic equipment and cabling that are required to control/readout a few dozen qubits already can fill up a whole rack~\cite{Krinner2019}. Despite recent progress in fabricating and packaging hundreds of qubits with high-yields on a single/multi-chip modules~\cite{das-2020,smcm-rigetti2021}, scaling up the interface electronics to support hundreds of qubits is reaching major roadblocks. This is mainly due to cooling power constraints in cryogenic systems, which limits the cabling heat load and, by extension, the wiring capacity. To address this challenge, researchers from Google and Intel have proposed miniaturizing essential control electronics to chip-scales using advanced cryogenic CMOS technologies~\cite{Patra2020,Bardin2019,Pauka2019}. However, due to the stringent cooling power constraint in cryogenic fridges ($1.5W$ at the \SI{4}{\kelvin} stage), pure electrical techniques would not be able to support more than $\sim$100 qubits in a dilution fridge~\cite{Krinner2019}. 

Fiber-based optical links are a potential solution for this problem, because optical fibers are made of significantly less thermally-conductive materials compared to RF coax cables. Moreover, the high-bandwidth of optical links can be exploited to transmit multiple control signals over a single fiber using wavelength-division multiplexing (WDM). Researchers from NIST have recently demonstrated such a fiber-based qubit control scheme by placing a photodiode with a SC qubit at the coldest stage of the dilution fridge: \SI{10}{\milli\kelvin} (MXC stage)~\cite{Lecocq2021}. However, this direct optical control of a SC qubit is not a scalable solution due to the active power dissipation from the optical-to-microwave conversion in the coldest stage of the dilution fridge, which also has the lowest cooling power of \SI{20}{\micro\watt}.

In this paper, we aim to study and compare the proposed approaches for controlling SC qubits in detail and present their limitations. We will propose an optimized cryogenic RF-photonic method to enable the XY control of thousands of SC qubits. Our scalable method exploits the advantages of optical fibers, while meeting all the power budget limitations of state-of-the-art dilution refrigerators. In the proposed method, RF control signals are generated at room temperature (RT) and transmitted to a cryogenic silicon photonic control chip at 4K through optical fibers. The cryogenic chip will perform the electro-optical conversion and drive the qubits using superconducting RF coaxial cables. Using optical links between the RT and 4K stages allows for the reduction of cabling complexity and heat load from the coaxial cables. Today, these cables are the main bottleneck for scaling up QCs. Our proposed architecture can outperform state-of-the-art and previously proposed cryo-CMOS~\cite{Patra2020,Bardin2019,Pauka2019} and optical~\cite{Lecocq2021,PRXQuantum.2.017002} solutions. 

Our analytical study in this work mainly focuses on the XY control interface rather than the readout path. Typically, the required number of cables (and signals) for readout of the qubit states is smaller than for the control of the qubit states (see Section~\ref{sec:implementations}), because frequency multiplexing can be used to share the readout lines among multiple SC qubits \cite{Heinsoo2018,Hornibrook2014}. However, presented approach in this paper can be also applied to the readout path as well. Additionally, our method can be efficiently implemented using cryogenic silicon photonic processes in the future.

This paper is organized as follows: Section~\ref{sec:background} briefly describes the principles of SC transmon qubits along with control and readout signal requirements. We will explain the thermal budget limitations of a cryogenic dilution fridge in Section~\ref{sec:thermallimits}. Section~\ref{sec:implementations} details the conventional and state-of-the-art implementations of the control interface in SC QCs. Section~\ref{sec:proposal} introduces our proposed RF-photonics approach and compares it to previously reported implementations. Finally, we summarize the results and discuss possible future directions based on our approach in Section~\ref{sec:discussion}.


%
%
%

\section{Background}
\label{sec:background}

\begin{figure}[t]
    \begin{center}
        \includegraphics[width=0.5\textwidth, keepaspectratio]{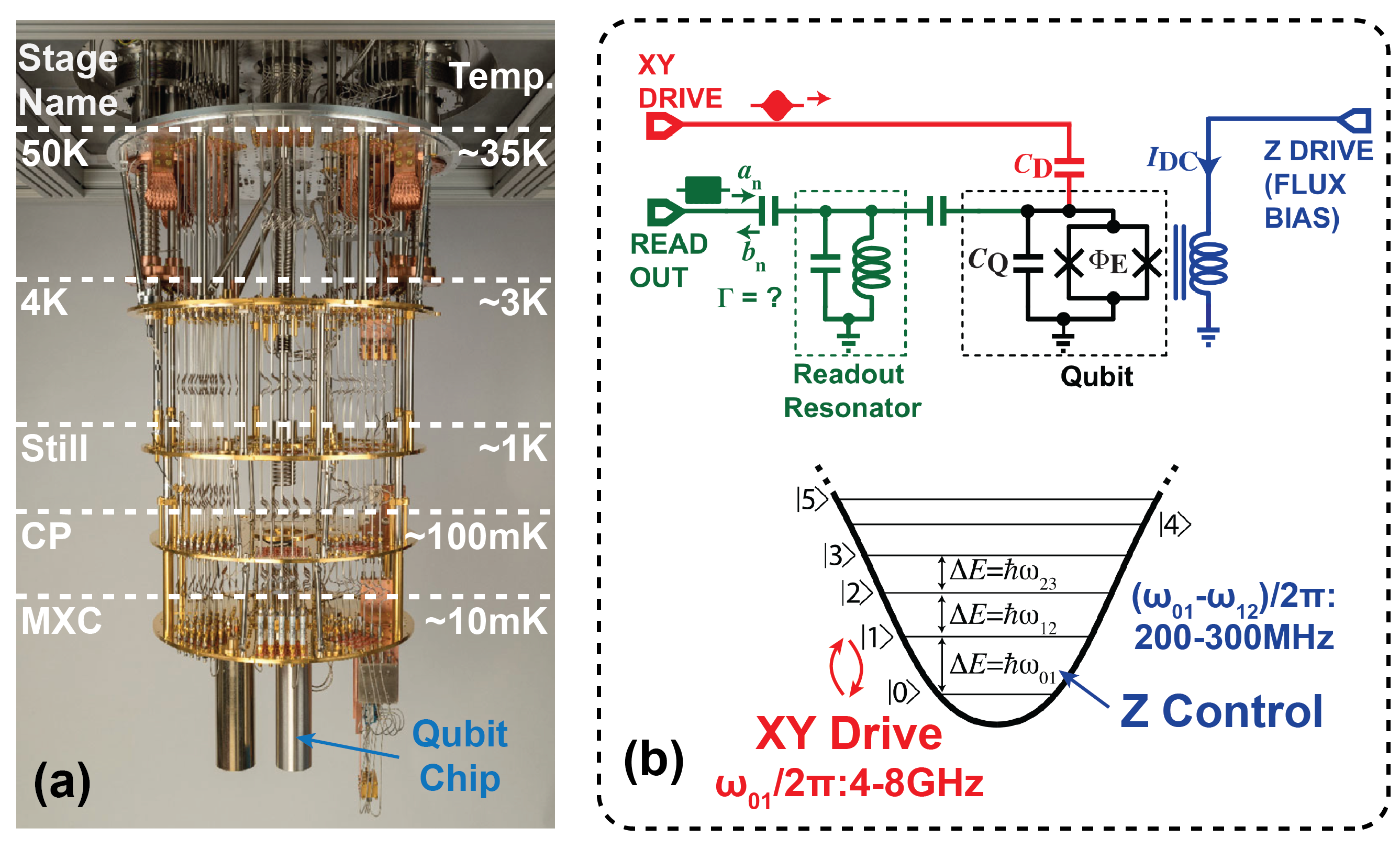}
        \caption
        {
           (a) An example dilution fridge (Bluefors XLD DR)~\cite{fridge-bluefors}, and (b) Schematic of a transmon qubit showing XYZ control ports and the energy diagram (from~\cite{bardin2021,bardin-isscc2022}).
        }
        \label{fig:transmonqubit}
    \end{center}
  \vspace*{-\baselineskip}
\end{figure}

In this section, we briefly describe the operation principles and RF signal requirements for transmon qubits. Transmon qubits are the most popular type of superconducting qubits made using Josephson junctions~\cite{blais2020quantum}. An example of a basic flux-tunable transmon qubit with complete control signals (XY and Z drive lines) and readout ports is shown in Fig.~\ref{fig:transmonqubit}b. The Josephson junction is a superconducting tunnel junction that acts as a non-linear inductance in this structure. This type of qubit consist of a capacitor ($C_Q$) in parallel with a Josephson junction loop, also known as a SQUID (superconducting quantum interference device), to form a frequency-tunable non-linear LC resonator~\cite{Bardin2019}. This non-linearity creates anharmonicity in energy levels as shown in Fig.~\ref{fig:transmonqubit}b diagram, allowing the transmon to be used as a qubit~\cite{bardin2021}. To operate the device in the quantum mechanical regime, the thermal noise level has to be suppressed well below the microwave photon temperature ($T_{Photon}$ is approximately $250mK$ for $5GHz$ resonance frequency). As a result, these qubits should operate in deep cryogenic temperatures ($10mK$ range).

\subsection{Transmon Qubit Control}

Quantum algorithms rely on a series of operations to control the states of individual qubits, generate entanglement between multiple qubits, and measure the states for readout. For transmon qubits, the control and readout is mainly carried out by microwave signals. Figure~\ref{fig:transmonqubit}b shows a basic schematic of a single transmon qubit. The Z drive control signal is used to initialize and control the qubit frequency by applying a current pulse of $10\mu A-1mA$ for $\sim$100ns duration~\cite{Majer2007}. To implement rotations in the XY plane of the Bloch sphere, the XY drive line is used. Microwave pulses at the resonant frequency of the qubit cause oscillations between the $\ket{0}$ and $\ket{1}$ states (corresponding to the first two energy levels). The wave envelope and the duration of microwave pulse affects the amount of rotation, while the carrier phase sets the axis of rotation. Hence, the XY-drive signal can be used to excite the state transitions. The excitation frequency between the first two energy states can be determined from $\omega_{01}=(E_1-E_0)/\hbar=\omega_0-E_C/\hbar$, where $E_C = q^2/2C_Q$ is the energy required to add one electron with a charge of $q$ to the capacitor charge~\cite{bardin2021}. We can also calculate the second transition energy as $\omega_{12}=(E_2-E_1)/\hbar=\omega_0-2E_C/\hbar$, and define the corresponding anharmonicity as $\eta = \omega_{12}-\omega_{01}=-E_C/\hbar=-q^2/{2\hbar C_Q}$. This metric determines the speed of XY gate operation and RF bandwidth requirements of the XY-drive RF signals~\cite{Bardin2019}.

The readout line is also used to perform the projective measurements of the qubit state via  a reflection or transmission of a linear readout resonator which is capacitively coupled to the qubit~\cite{Bardin2019}. However, wiring the microwave readout signals is less critical than control signals since the readout paths can typically be shared among multiple qubits using frequency multiplexing~\cite{bardin-isscc2022,Arute2019}. Thus, we will focus on the control signals for the rest of discussions in this paper.

\subsection{Signal Requirements}

Typical transmon qubits have transition frequencies ($\omega_{01}$) in the range of $4-8GHz$. In this paper, we consider the RF signal carrier frequency of \SI{6}{GHz} since typical control and readout frequencies are centered around that. The actual required signal power to control a qubit ($P_Q$) is small, on the order of \SI{-70}{dBm} (e.g., a peak power of \SI{-66}{dBm} for a \SI{20}{ns} long $\pi$-pulse)~\cite{Bardin2019}). Additionally, the spectral width of XY-drive microwave excitation pulses are bound by the anharmonicity ($\eta /{2\pi}$), which is typically in the \SI{200}{MHz}–\SI{300}{MHz} range. The suppression ratio between the $\omega_{01}$ and $\omega_{12}$ transitions should be $\sim50dB$ for acceptable gate fidelities~\cite{Bardin2019}. Therefore, pulse shaping techniques are used on the control pulse to minimize decoherence while preventing off-resonant transitions for high gate fidelity. These pulses have durations between \SI{10}{ns} and \SI{30}{ns}~\cite{bardin2021}.

The thermal photon occupation has to be suppressed to only a few $10^{–3}$ at the qubit level~\cite{Krinner2019}. This requires the thermal noise power spectral density to be below \SI{-205}{dBm/Hz}, which is equivalent to a thermal noise current spectral density of $\sim2pA/\sqrt{Hz}$~\cite{Krinner2019}.

Conventionally, control RF signals and pulses for XY/Z-drive lines are generated via commercial rack-mount RF arbitrary waveform generators (AWG) at RT as shown in Fig.~\ref{fig:imp}a. These AWGs typically operate at $1GS/s$ with 14-bit accuracy to generate the base-band signal that can be up/down converted via single sideband (SSB) mixers with LO and RF signals in the microwave carrier frequency range of ($4GHz-8GHz$)~\cite{Bardin2019}. The Z-flux line pulses typically only require $0.5GHz-1GHz$ bandwidths, and thus, the Z-control and XY-control can be combined together at the qubit stage using a bias tee~\cite{bardin-isscc2022, maneti-apl2021} to minimize the wiring to a qubit chip. To achieve target noise levels, a total attenuation of more than $60dB$ and $20dB$ for XY and Z lines, respectively, is typically required to reduce black-body radiation present in cables at RT~\cite{Krinner2019}. Since the bandwidths of Z-lines is much smaller than the qubit frequency, the thermal noise at the qubit frequency can be suppressed by using a low-pass filter before combining the XY and Z lines. Overall, required RF cabling of these signals ultimately limits the number of qubits that can be operated in today's dilution fridges as discussed in Section~\ref{sec:thermallimits}. 

While we are only focusing on the RF signal delivery for the XY drive lines in this paper, we note that RF cabling of the Z control lines is also another critical limit in today's systems. This is due to the fact that Z-lines also require RF coaxial cables to maintain the signal integrity of flux-control pulses, and majority of cables in the latest demonstrations (e.g., by Google~\cite{bardin-isscc2022}) have been used to deliver these signals. 

\section{Thermal Budget Limitations}
\label{sec:thermallimits}

Transmon qubits have to be cooled down to deep-cryogenic temperatures in a dilution fridge as shown in Fig.~\ref{fig:transmonqubit}a. These fridges normally comprise of multiple stages ranging from RT down to MXC, where the qubit chip resides. Thermalization of RF and DC cables, attenuators, and other microwave components at any of these stages limits the overall number of qubits. Hence it is crucial to reduce the heat load as well as thermal radiation to the qubits. In this section, we will introduce the thermal limitations of dilution fridge stages and the various sources that contribute to the heat load.

\subsection{Cooling Power}
Each stage of a dilution refrigerator has a limited capacity to extract the generated heat load to maintain the target stage temperature. This  maximum amount of heat load that each stage can handle is called the cooling power. Hence, it is crucial for each stage to have a total heat load below its cooling power. Table~\ref{tab:coolingpower} shows the list of typical cooling power for a state-of-the-art dilution fridge~\cite{fridge-bluefors}; the cooler stages in the dilution fridge have smaller cooling powers. This information can be used to make informed decisions about which stages to use and how to spread the system and microwave components across them, while maximizing the number of qubits that the system can support.

\begin{table}[!htp]
    \centering
    \caption{Stage names, temperatures and cooling power for Bluefors XLD400 DR}
    \begin{tabular}{|c|c|c|}
        \hline
          Stage Name    & Typical Temperature & Cooling Power(W)\\ \hline
          50 K          &35K       & 30   \\ \hline
          4 K           & 2.85K       & 1.5 \\ \hline
         Still          &0.882K      & $40  \times 10^{-3}$  \\ \hline
              CP       &0.082K      & $200 \times 10^{-6}$ \\ \hline
          MXC           &0.006K         & $19 \times  10^{-6}$ \\ \hline
         
    \end{tabular}
    \label{tab:coolingpower}
 \vspace*{-\baselineskip}
\end{table}

\subsection{Sources of Heat Load}
Heat load refers to the amount of thermal heat that is leaked into or generated by the components and chips in each stage. This includes heat flow from higher temperature stages to lower temperature stages due to the coaxial cables, and microwave components~\cite{Krinner2019}. Notice that heat loads related to the fridge itself, such as heat leakage through the post and mechanical body, is already considered in the calculation of the total cooling power of stages; therefore, we can ignore them in our modeling. Total heat load can be simply calculated as the sum of passive and active heat loads that are described below:

\subsubsection{Passive Load}
In a dilution fridge setup, RF control signals are typically generated at RT and sent from one stage to the next stage using RF coax cables that run between these stages. The heat flow from a higher temperature stage to a lower temperature stage is the main source of passive heat load for the lower temperature stage. This heat load depends on the cable material and diameter, and the temperature of each stage.

One example of cables that are commonly used to send microwave signals are stainless steel RF coaxial cables. These cables normally have a passive heat load of around \SI{1}{mW} per cable for the 4K stage~\cite{Krinner2019}. It is important to note that wiring a signal between two cryogenic stages (with temperatures below 4K), we can exploit the superconductivity of materials such as niobium-titanium (NbTi) or copper-nickel (CuNi) to reduce the cable diameters and effective heat conductances. These SC coax cables can drastically reduce the passive heat load down to the $\mu W-nW$ range for Still, CP, and MXC stages~\cite{Krinner2019}.


Another type of cable we are proposing to use in this work is the fiber optic cable; the ultra-low thermal conductivity of the fiber reduces the passive heat load. Specially, single-mode optical fibers, made out of silica, have a total diameter of only \SI{250}{\micro\meter} and, recently, it has been utilized to interface with SC qubits~\cite{Lecocq2021}. Additionally, the higher bandwidths of optical interconnects can be used to reduce the total number of fibers required to control a large number of qubits within the thermal limits of a dilution fridge. We will elaborate on these benefits in Section~\ref{sec:discussion}. However, we will first estimate the passive heat load of single-mode fibers below which is critical for the rest of discussions in this paper. 

\textit{Passive Heat Load Estimation for an Optical Fiber:}
We estimated the passive heat load of the optical fiber connected between RT and the 4K stages for a single-mode optical fiber (SMF28), as will be used in our analysis. The heat flow, $P$, between $T_1 = 4K$ and $T_2 = RT$ is calculated using the below equation based on Fourier's Law: 
\begin{equation}
    \label{eq:ofhf}
    P = \int_{T_1}^{T_2} \frac{\rho_b(T)A_b + \rho_{cl}(T)A_cl + \rho_c(T)A_c}{L} dT,
\end{equation}
where the thermal conductivities and cross sections are defined as $\rho_b, \rho_{cl}, \rho_c,$ and $A_b, A_{cl}, A_c$ for the buffer, cladding, and core, respectively, of the optical fiber. $L$ is the length of fiber, of which we approximate that to be around 1 meter. Silica ($SiO_2$) is used for the material of the core and the cladding, while polymers, including Teflon, are used for the buffer material~\cite{weik2012fiber}. The thermal conductivities of $SiO_2$ and Teflon were determined using the respective data from~\cite{marquardt2002} and~\cite{simon1994}. Using Equation ( \ref{eq:ofhf}), the passive heat load of the optical fiber from RT to 4K was calculated to be $\sim5.6\mu W$. 

\subsubsection{Active Load}
The active load generally depends on the RF signal power losses (Joule heating either in the coax cables or attenuators) and the power dissipated by the active components such as microwave high-electron-mobility transistor (HEMT) amplifiers (used for readout). There are three different contributors to active load on the dilution fridge that we consider in this work. First, any attenuation added along the cables for thermal noise suppression, as discussed in Section~\ref{sec:background}, will contribute to the heat load of the stage it occurs in. Second, the active heat load contributed by the dissipation of applied microwave signals in the RF cables. These two effects can be modeled together by adding the total loss of attenuation and the cables. Third, any active component, including the HEMT amplifier, cryogenic-CMOS chip or photonic devices which dissipates power, will contribute to the active head load. This category also includes dissipation due to DC signals for biasing components, such as HEMT amplifiers at the 4K stage or flux bias qubits at MXC~\cite{Krinner2019}.

Notably, the electro-optical conversions required for interfacing with qubits using optical interconnects can lead to excess active heat loads. In particular, the conversion of high-energy optical photons to microwave photons through a photodiode at cryogenic stages (similar to~\cite{Lecocq2021}) causes almost the entire optical power to be fully dissipated as heat. We will discuss this critical note later in the following sections.


\section{State-of-the-art Methods of Controlling Transmon Qubits}
\label{sec:implementations}

In this section, we will introduce three major demonstrated approaches to interfacing with multiple SC transmon qubits. We will focus on the scalability aspect of these methods in terms of the thermal budget limitations of dilution fridges. We will compare these limitations and both the active and passive power per qubit in with our proposed solution in Section~\ref{sec:discussion}.

\begin{figure}[t]
    \begin{center}
        \includegraphics[width=0.5\textwidth]{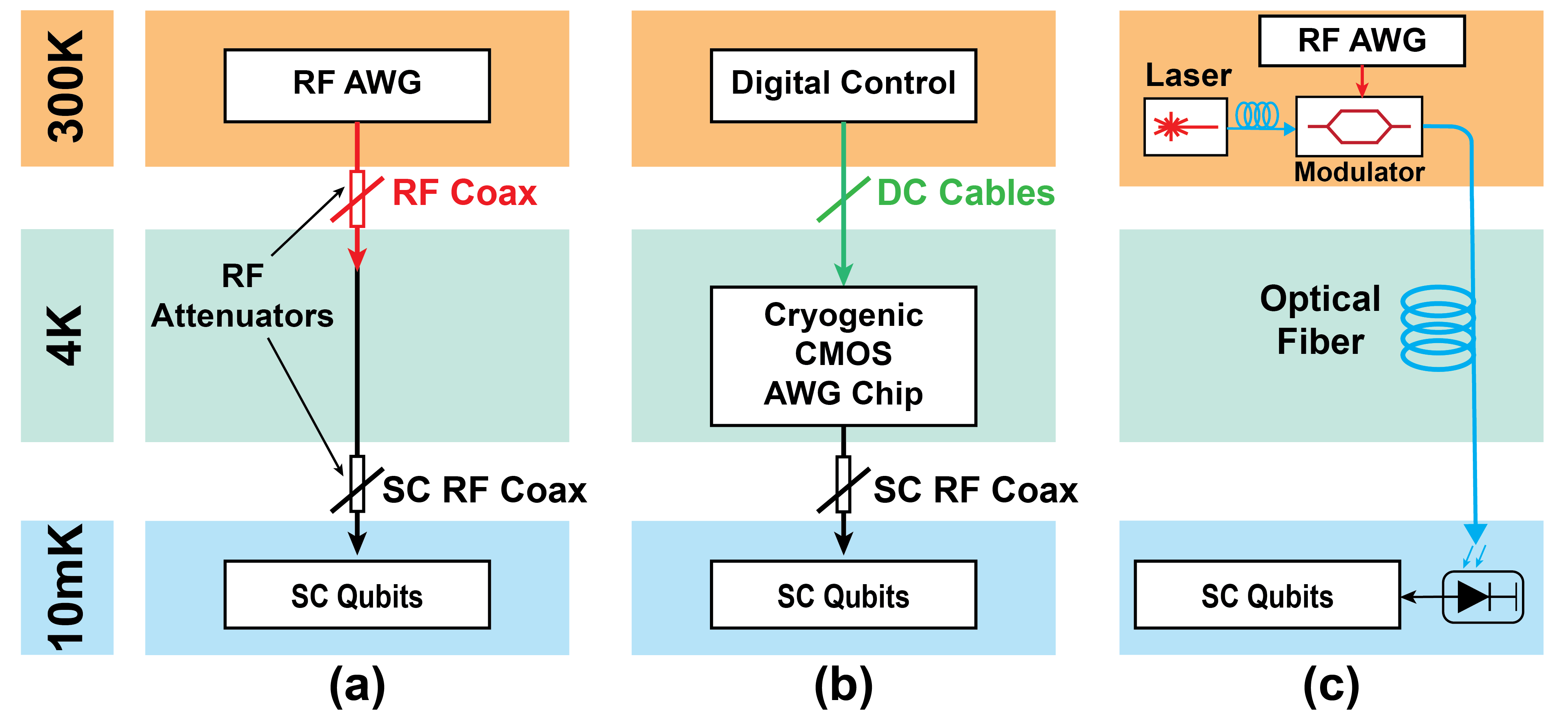}
        \caption
        {
           Proposed methods to control a transmon qubit: (a) Conventional approach using RT equipment, (b) Cryogenic CMOS approach, and (c) Deep-cryogenic RF-photonic approach.
        }
        \label{fig:imp}
    \end{center}
    \vspace*{-\baselineskip}
\end{figure}

\subsection{Conventional RF Coax}
\label{sec:conventional}

The conventional method of interfacing with transmon qubits is using a RT electronics control and measurement system (Fig.~\ref{fig:imp}a). In this method, RF signals (see Section~\ref{sec:background}) are generated and measured using rack-mounted commercial AWG or FPGA platforms~\cite{Xu2021QubiCAO} along with RF I/Q mixers~\cite{Xu2021RadioFM}. Despite recent demonstrations of QCs with over 100 qubits using this method~\cite{Arute2019,huang2020superconducting}, the physical size, cost and electrical failure rate have become the limiting factors when considering the scalability of this approach~\cite{Xu2021RadioFM}. More importantly, because thermal photons propagate through the coax cables towards the lower cryogenic stages, it is crucial to reduce the spectral density of thermal radiation by installing a series of attenuators in the microwave coax lines~\cite{Krinner2019}. This adds additional active heat load to the stages in addition to the passive heat load of cables that was already considered. 

Briefly, we will summarize the calculation for the required RF attenuation (based on~\cite{Krinner2019}) as it will be adopted to analyze and optimize our proposed solution in Section~\ref{sec:proposal}. The two-sided power spectral density (PSD) of thermal voltage noise for a resistor load or RF coax cable with characteristic impedance of $R$ can be calculated using:

\begin{equation}
    \label{eq:vthermnoise}
    S^{th}_{v}(T, \omega) = 2R\hbar \omega n_{BE}(T, \omega).
\end{equation}

$n_{BE}(T, \omega)$ denotes the Bose-Einstein distribution ($1/[exp(\hbar \omega /k_{b}T) - 1]$) for photon occupancy. This term determines the photon flux spectral density or the number of photons per Hz frequency interval per second. If $\hbar \omega \ll k_B T$, the noise voltage simplifies to classical Johnson–Nyquist noise, $S_{V}^{th}(T) = 2 k_B T R$ ($k_B$ is the Boltzmann constant). The voltage noise can be converted to current noise, $S^{th}_{I}(T, \omega) =  S^{th}_{v}(T, \omega)/R^2 $. As the thermal photon occupation at the MXC stage should be smaller than $10^{-3}$, we can estimate the total attenuation required for the RF signals that are generated at RT from:

\begin{equation}
    \label{eq:atten}
    \frac{n_{BE} (300K, 6 GHz)}{n_{BE}(10mK, 6GHz)} = \frac{n_{BE} (300K, 6 GHz)}{10^{-3}} \approx 60 dB
\end{equation}

For a more accurate estimation, we need to note that the distribution of the total attenuation among the stages can impact the actual thermal noise at the qubit, because each stage has a different temperature. Additionally, optimizing the distribution of attenuation is crucial for reducing the active heat load overhead for the stages. In order to figure out the optimal attenuation (denoted by $A_i$) at stage $i$, we can use the following equation:

\begin{equation}
    \label{eq:noisepstage}
     n_i (w) = \frac{n_{i-1}(\omega)}{A_i} + \frac{A_i - 1}{A_i} n_{BE}(T_{i,att}, \omega),
\end{equation}

where $n_{i-1}(\omega)$ is the noise photon occupation number of the $i$-th stage with a temperature of $T_{i,att}$. By writing Equation (\ref{eq:noisepstage}) for each of the stages, we can find the optimal placement of attenuators that achieves the thermal noise target at the sample. It has been shown that placing $\sim$20dB attenuation at each of the 4K, CP, and MXC stages can be optimal~\cite{Krinner2019,bardin2021}. This implementation leads to the estimate that ultimately $\sim$150 qubits can operate within the thermal budget of a state-of-the-art fridge~\cite{Krinner2019}. Currently, the main limitations are the active heat load of the MXC stage (due to the attenuators). Ultimately, the upper bound of the number of cables in this approach is limited to $\sim$1000, due to the passive heat load of RF coax cables at 4K stage. For example, Google's Sycamore quantum processor with 54 transmon qubits required over 200 long coaxial cables: 54 cables for XY control, 54 cables for Z control, 88 cables to control two-qubit gates, and about 36 cables for the readout~\cite{bardin-isscc2022}.

\subsection{Cryogenic CMOS}
\label{sec:Cryocmos}

The passive heat load limitations of the 4K stage in the conventional approach and the massive complexity of cabling have recently motivated researchers to design custom ultra-low power CMOS chips that can generate required RF signals at the 4K cryogenic stage (Fig.~\ref{fig:imp}b). In doing so, the cabling can be significantly simplified from RT to 4K stage by running only a few DC and low-speed digital I/O wires to the cryogenic CMOS chip. These wires have about two orders of magnitude lower passive heat load compared with stainless-steel RF coax~\cite{Krinner2019}. This approach has been pursued by Google~\cite{Bardin2019}, IBM~\cite{frank-isscc2022}, and Intel Horseridge chips (for the semiconductor-based spin qubit platforms)~\cite{park-isscc2021}. However, the active heat load due to the CMOS chip power in this method has been dominating the 4K stage heat load. The lowest reported power is around \SI{2}{mW} per qubit using a \SI{28}{nm} technology node~\cite{Bardin2019}, while not yet achieving acceptable gate fidelity compared with the conventional method.
Additionally, the excess noise of CMOS electronics still has to be suppressed via microwave attenuators ($>40dB$) before reaching the qubit; therefore, this approach does not mitigate the active heat load issue for stages below 4K.

\subsection{Deep-cryogenic Photonic}
\label{sec:Cryophotonics}

Another approach to overcome the passive heat load of the 4K stage is to use optical interconnects to deliver the RF signal to the qubits (RF-photonic links). As mentioned in Section~\ref{sec:thermallimits}, optical fibers have orders of magnitude less passive heat loading compared with RF coax. In this approach, RF control signals are generated at RT and imprinted on a laser carrier (e.g. 1310nm or 1550nm) via an optical modulator (e.g. Mach-Zehnder or ring modulator). Furthermore, ultrahigh optical spectrum band can be used to communicate multiple control/readout RF signals through the same fiber using wavelength-division multiplexing (WDM).

Recently, this approach has been implemented~\cite{Lecocq2021} by placing a photodiode in the MXC stage, as illustrated in Fig.~\ref{fig:imp}c, and using the RF photo-current to control and readout a transmon qubit~\cite{Lecocq2021}. While this scheme simplifies the wiring by potentially eliminating the need for any coax cables, it has a major disadvantage of increasing the active heat load in the MXC stage, which has the most stringent cooling power budget ($\sim 20\mu W$). The active heat load overhead will be in the range of $\sim 1\mu W$ that can be calculated from the required signal power at the qubit, and the corresponding optical power level at the photodiode. In practice, there will be also other power overheads, including dark current of photodiodes, and the tuning power for multiple resonant photonic elements such as micro-ring resonators (in case of WDM).



\section{Proposed Cryogenic RF-photonic Approach}
\label{sec:proposal}
\begin{figure}[t]
    \begin{center}
        \includegraphics[width=0.45\textwidth, keepaspectratio]{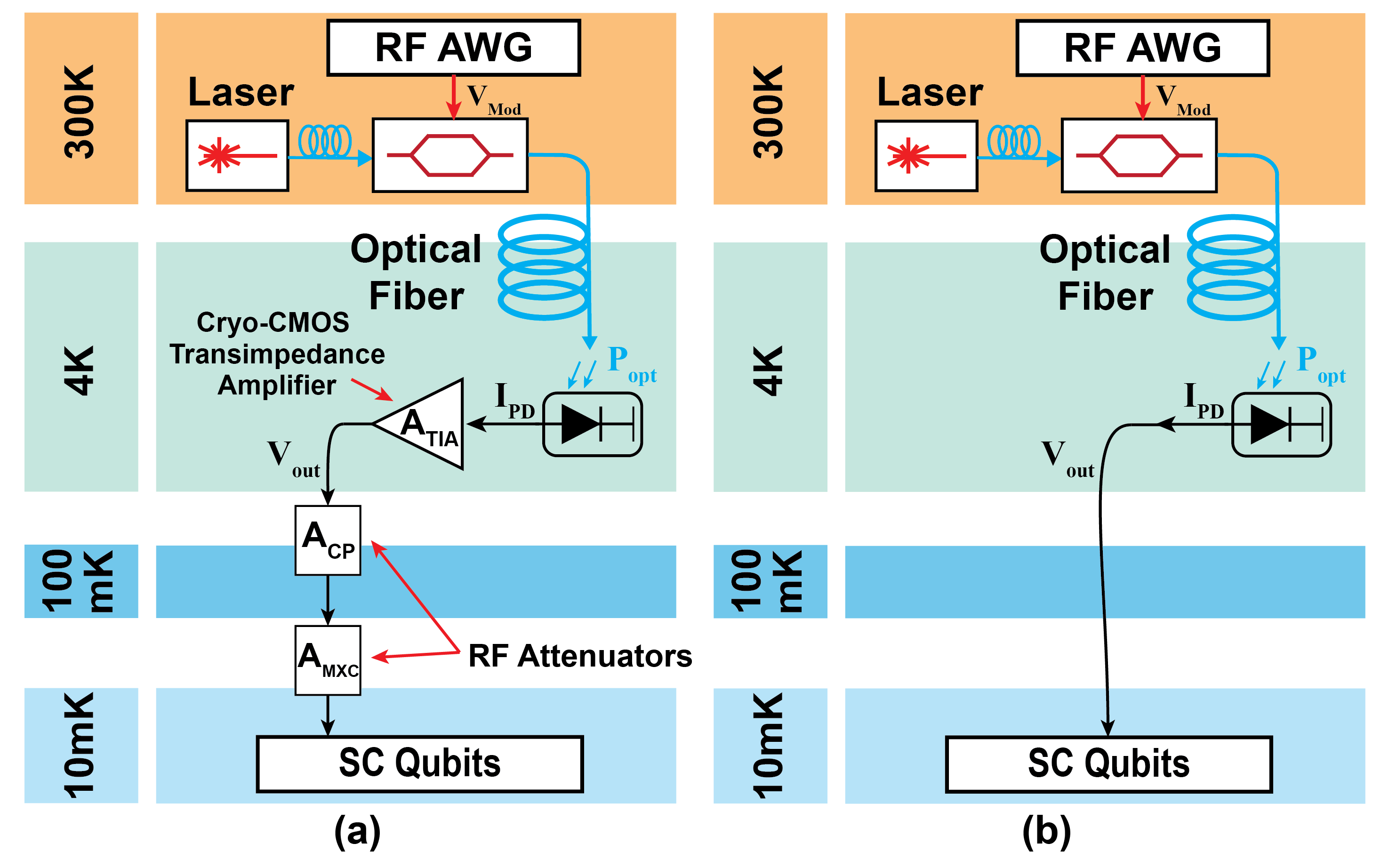}
        \caption
        {
           Proposed approach based on cryogenic RF-photonics: (a) Generic case for system-level modeling, and (b) Simplified case with no cryogenic amplifier.
        }
        \label{fig:proposed}
    \end{center}
    \vspace*{-\baselineskip}
\end{figure}

We are proposing a modified RF-photonic solution, depicted in Fig.~\ref{fig:proposed}, that can overcome the scalability limits of state-of-the-art methods as described in Section~\ref{sec:implementations}. While we are proposing the use of optical RF links, we argue that all the electro-optical conversions should occur at the 4K stage as opposed to the MXC stage (deep-cryogenic method~\cite{Lecocq2021}) for scalability. For this argument, our methodology is to first calculate the overall qubit noise, and estimate the passive and active heat loads for each stage. To do so, we model the proposed system in the most generic case as shown in Fig.~\ref{fig:proposed}a. In this model RF-photonic signals are converted back to the electrical domain at the 4K stage, amplified via a cryogenic CMOS low noise trans-impedance amplifier (TIA), and transmitted to the MXC stage through SC coax cables. We have also added potential RF attenuators at the CP and MXC stages. Notice that RF cables connecting the cryogenic stages are SC wires with minimal heat transfer as opposed to RT coax cables. This point will be discussed in Section~\ref{sec:discussion}).

\subsection{Noise Analysis}
\label{sec:noisesource}

We propose the use of optical fibers to transmit the signals between the RT to 4K stage, eliminating the thermal noise of RT coax cables. However, we need to model and consider the thermal noise for cryogenic stages, and include all other noise sources that can be coupled into the qubit signal as listed below. For noise modeling, we use one-sided PSDs in this section.

\subsubsection{Shot Noise}
\label{sec:shotnoise}
Photo-detection at the 4K stage fundamentally adds shot noise into the photocurrent ($I_{PD}$), which can limit the qubit gate fidelity~\cite{Lecocq2021}. The PSD of shot noise current is given by $S_I^{Shot}(\omega) = 2q\overline{I}|H(\omega)|^{2} $, where $\overline{I}$ is the average photocurrent and $H(\omega)$ is the transfer function of photodiode. We assumed $H(\omega) = 1$, as the cryogenic photodiodes can achieve $>10GHz$ bandwidths~\cite{Lecocq2021}. Shot noise will be $\sim 0.6 pA/\sqrt{Hz}$, for an average current of $1 \mu A$.

\subsubsection{Electro-optical Modulator Noise}
\label{sec:eomnoise}
The thermal voltage noise of the drive signal ($S^{V_{dr}}_{v}(\omega)$) that is generated at RT for the electro-optical modulator (EOM) will be propagated to the qubit via optical link. We can write the optical output of the modulator simply in the form of $ P_{opt} = \overline{\rm P_{Laser}}(1+\cos(\pi v(t)/v_{\pi}))$, where the cosine term models the voltage-controlled phase shift effect in a Mach-Zehnder modulator. As a result, the PSD of photocurrent noise can be derived from $S^{EOM}_I(\omega) = S^{V_{dr}}_{v}(\omega) \cdot (\overline{\rm I} \pi/(V_{\pi}))^2$~\cite{Lecocq2021}. We can estimate $S^{V_{dr}}_{v}(\omega) = 4k_BTZ_{dr}$ with $T=300K$ and $Z_{dr}=50 \ohm$. For simplicity, we ignored the excess noise of the driver circuitry. For $\overline{\rm I} = 1 \mu A$, and $v_{\pi}=2V$, we can estimate that $S^{EOM}_I(\omega) \approx 1.5 fA/\sqrt{Hz}$. 


\subsubsection{Laser RIN Noise}
\label{sec:rin}
The laser's optical power fluctuations known as relative intensity noise (RIN) noise can also impact the qubit signal integrity. The current noise PSD can be found using $S_I^{RIN}(\omega) = \overline{\rm I}^{2} \cdot RIN(\omega)$~\cite{Lecocq2021}. Majority of high-quality lasers have an $RIN(\omega)$ of below $-150dB/Hz$~\cite{fatadin2006numerical}, which leads to $S_I^{RIN}(\omega) \approx 30 fA/\sqrt{Hz}$ for $\overline{\rm I} = 1 \mu A$.

\subsubsection{Laser Phase Noise}
Laser phase noise must be also considered in the overall noise calculation. As we are only proposing an amplitude modulation of the RF signal in our optical links, we can ignore this source of noise. However, for more complex and coherent modulation schemes in future, this noise source need to be considered~\cite{Yan2021}.

\subsubsection{Thermal Phase Noise of the Fiber}
The thermal phase noise of the fiber or the optical Nyquist noise is the thermal noise or fluctuation in the reference index and phase~\cite{fiberthermnoise}. This noise source is negligible for fibers with only a few meter-long length. Similar to the phase noise, this source of noise can be ignored in our direct detection approach.

\subsubsection{Amplifier's Noise}
A low-noise and low-power TIA can be also used after the photodiode to provide an extra gain ($V_{out}/I_{PD}$) of $A(\omega)$. The TIA can reduce the required photocurrent at the cost of extra active power load ($P_{TIA}$) and added noise. To model this added noise in our model, we assume that the TIA has a noise figure (NF) that can be defied as $NF = 1 + T_e/T_0$, where $T_e$ and $T_0$ are the noise and ambient temperatures, respectively. Hence, the PSD of the input current noise of the TIA can be written as:

 \begin{equation}
    \label{eq:cmosnoise}
    S^{TIA_{in,eq}}_{I}(\omega) = (NF - 1) \times 4k_BT_0/Z
\end{equation}

with $Z=50\ohm$ and $T_0=4K$ equal to the stage temperature. Recently, low noise amplifiers (LNA) have been reported with an NF below \SI{1}{dB}~\cite{patra-jssc2017,wong-ims2020}, and we have considered this range in our analysis. While we have included the TIA noise in our model for the completeness and analyzing the noise trad-offs, avoiding the TIA is preferred due to its power overhead. This is due to the fact that power consumption of these LNAs is still in the $mW$-range, which can dominate the active heat load for the 4K stage. In the simple case where no amplifier has been used (Fig.~\ref{fig:proposed}b), we can simply set $NF=0dB$, and $A_{TIA}=50\ohm$. In this scenario, there will be no active load overhead ($P_{TIA}=0$).

\subsection{Noise Estimation}
\label{sec:xy-noise}
Our goal is to determine the required optical power, attenuation, and TIA gain such that we can achieve the target noise and signal levels (see Section~\ref{sec:background}) at the MXC stage for qubit. Therefore, we first calculate the total PSD of noise at the MXC stage ($S_I^{MXC}$). Using the major noise sources described so far, the photodiode's current noise is calculated to be:

\begin{equation}
    \begin{aligned}
         S^{I_{PD}}_{I}(\omega) = S^{Shot}_{I}(\omega) + S^{RIN}_{I}(\omega) + S^{EOM}_{I}(\omega) \\
     = 2q\overline{\rm I} + (RIN(\omega) + 4k_BTZ \times (\frac{\pi}{v_{\pi}})^2)\times \overline{\rm I}^2
    \end{aligned}
     \label{eq:phcurr}
\end{equation}

Using the form of Equation~(\ref{eq:noisepstage}) for the CP and MXC stages, we can derive the current noise for all the stages as follows:

\begin{equation}
    \label{eq:4knoise}
     S_I^{4K}(\omega) = (S^{I_{PD}}_{I}(\omega) + S^{TIA}_{I}(\omega)) \times \abs{\frac{A(\omega)}{Z}}^2
\end{equation}

 \begin{equation}
    \label{eq:cpnoise}
     S_I^{CP}(\omega) = \frac{S_{I}^{4K}(\omega)}{A_{cp}} + \frac{A_{cp}-1}{A_{cp}}\cdot S_{I}^{th}(100mK, \omega)
\end{equation}

 \begin{equation}
    \label{eq:mxcnoise}
     S_I^{MXC}(\omega) = \frac{S_{I}^{CP}(\omega)}{A_{MXC}} + \frac{A_{MXC}-1}{A_{MXC}}\cdot S_{I}^{th}(10mK, \omega)
\end{equation}

In all the equations, $Z=50\ohm$, which is also assumed to be the characteristic impedance of coax cables. Additionally, we can calculate the signal power at the qubit level using:

\begin{equation}
    \label{eq:pqmin}
     P_{Q}(\omega) = \frac{(\overline{\rm I}\times A(\omega))^2}{Z \times A_{CP}A_{MXC}}
\end{equation}

$P_Q$ should be $\approx -70dBm$ for $\omega/2\pi=6GHz$. For the first order estimation, we can assume that the 4K stage noise will be the dominant factor. Hence, we can derive the qubit noise by combining Equations (\ref{eq:cpnoise}) and (\ref{eq:mxcnoise}):

 \begin{equation}
    \label{eq:attenuation}
     S_I^{MXC}(\omega) \approx \frac{S_I^{4K}(\omega)}{A_{CP} \cdot A_{MXC}} 
\end{equation}

Now we can use Equations~(\ref{eq:4knoise}) and~(\ref{eq:pqmin}) to obtain the MXC stage noise as: 

 \begin{equation}
    \label{eq:attenuation1}
     S_I^{MXC}(\omega) \approx \frac{P_{Q}}{Z \cdot \overline{\rm I}^2} \times (2q\overline{\rm I } + 4k_BT_{RT}Z\frac{\pi}{v_{\pi}}\overline{\rm I}^2 + (NF-1)\frac{4k_BT_{4K}}{Z})
\end{equation}
where $T_{RT}=300K$ and $T_{4K}=4K$. By setting $P_Q=-70dBm$, we can plot the PSD of current noise at the MXC stage versus the average photocurrent ($\overline{\rm I}$) for various NF values as shown in Fig.~\ref{fig:imp}. The dashed line in this plot shows the maximum tolerable qubit noise level. 
While larger $\overline{\rm I}$ leads to lower noise, it will also increase the active heat load in the 4K stage, which is discussed in the next subsection. Thus, an optimal design choice of the system will be the smallest $\overline{\rm I}$ as long as $A(\omega)/A_{CP}A_{MXC}$ is within a reasonable range given the active heat load limits. Ideally, we can find the required $\overline{\rm I}$ and $A(\omega)$ under the assumption that no amplifier or attenuator is needed ($A_{CP}A_{MXC} = 1$ and $A(\omega)= 50\ohm$). This leads to $\overline{\rm I} \approx 1.4\mu A$, achieving the noise level of $S_I^{MXC} < 0.7pA/\sqrt{Hz}$. We will use this case for the calculation of heat loads in the next section. However, the development of cryogenic TIAs with $\mu W$-level power consumption can make this scenario sub-optimal.  

\begin{figure}[t]
    \begin{center}
        \includegraphics[width=0.5\textwidth, keepaspectratio]{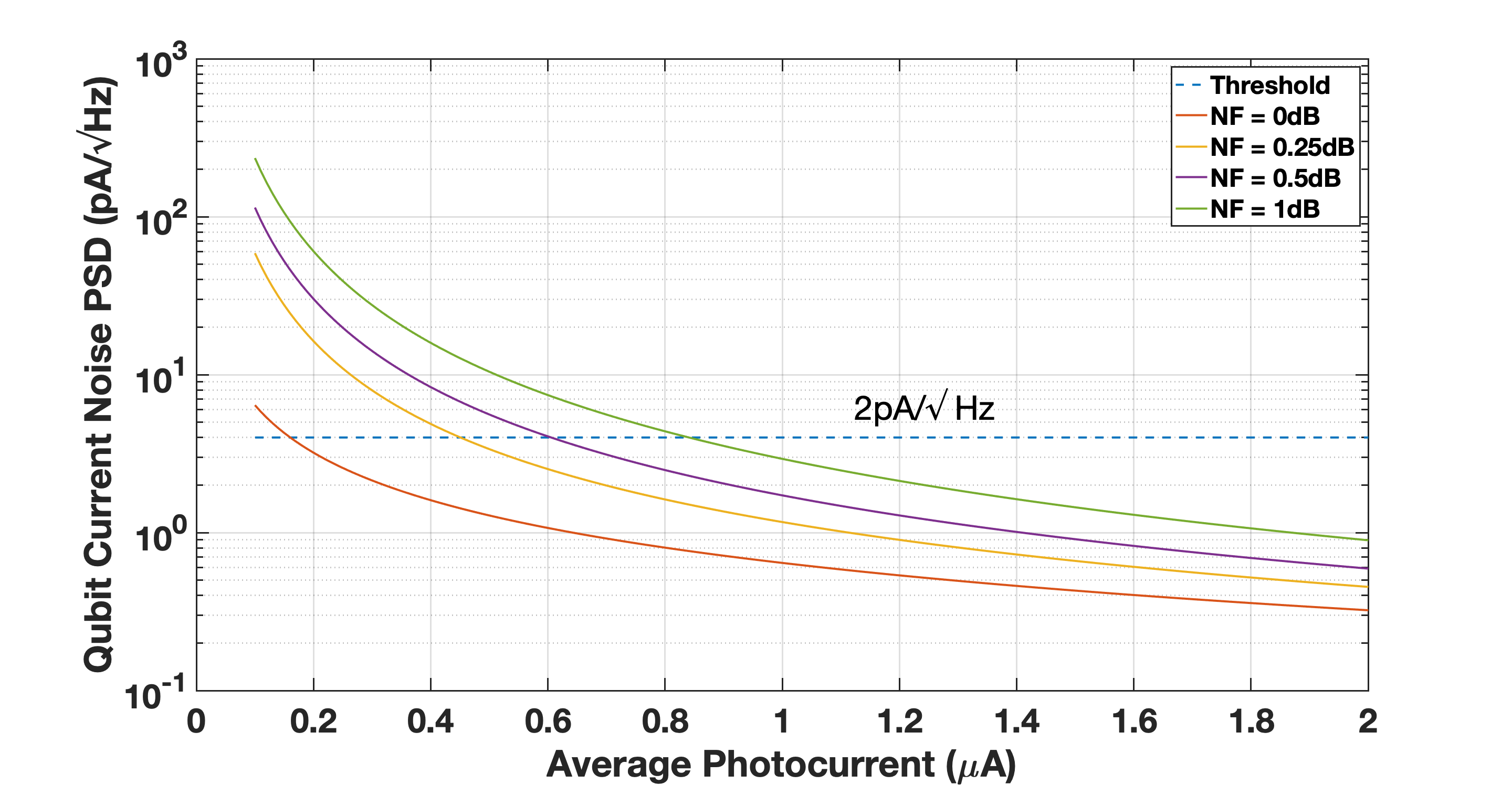}
        \caption
        {
           The PSD of qubit current noise ($S_I^{MXC}$) vs. average photocurrent ($\overline{\rm I}$) for various amplifier NF values. 
        }
        \label{fig:noisecurrml}
    \end{center}
      \vspace*{-\baselineskip}
\end{figure}

\subsection{Heat Load Estimation}
Table~\ref{tab:heatload} summarizes the overall critical heat load parameters for the proposed approach. The passive heat load of the CP and MXC stages is determined using the measurements from of~\cite{Krinner2019} adjusted for the 0.034" type NbTi-NbTi superconducting RF coaxial cables from KEYCOM~\cite{keycom-sccoax}. We have formulated the active heat load of each stage for the general case (Fig.~\ref{fig:proposed}a). For the case where no amplifier or attenuator is implemented, the only active heat load will be from the optical power at the 4K stage that can be derived from $\overline{\rm I} / PD_{Res}$, where $PD_{Res}$ will be the photodiode's responsivity that should be $\sim1A/W$~\cite{Lecocq2021}.

\begin{table}[bp]
    \centering
    \caption{Heat Load Parameters for Proposed Approach} 
    \begin{tabular}{|c|c|c|}
        \hline
         Stage & Passive Heat Load & Active Heat Load\\ \hline
         4K          & One fiber $\approx$ $5.6 \mu W$         & $P_{opt} = \overline{\rm I} / PD_{Res}$ \\ 
                     &                                & $P_{TIA} = 0$ \\ \hline
         CP       & One SC coax $\approx$ 0.06$\mu$W   & $(A_{CP} -1)\times A_{MXC}P_Q = 0$      \\ \hline
         MXC        & One SC coax $\approx$ 0.004$\mu$W & $(A_{MXC} - 1)P_Q = 0$ \\ \hline
    \end{tabular}
    \label{tab:heatload}
\end{table}

\begin{figure}[t]
    \begin{center}
        \includegraphics[width=0.45\textwidth, keepaspectratio]{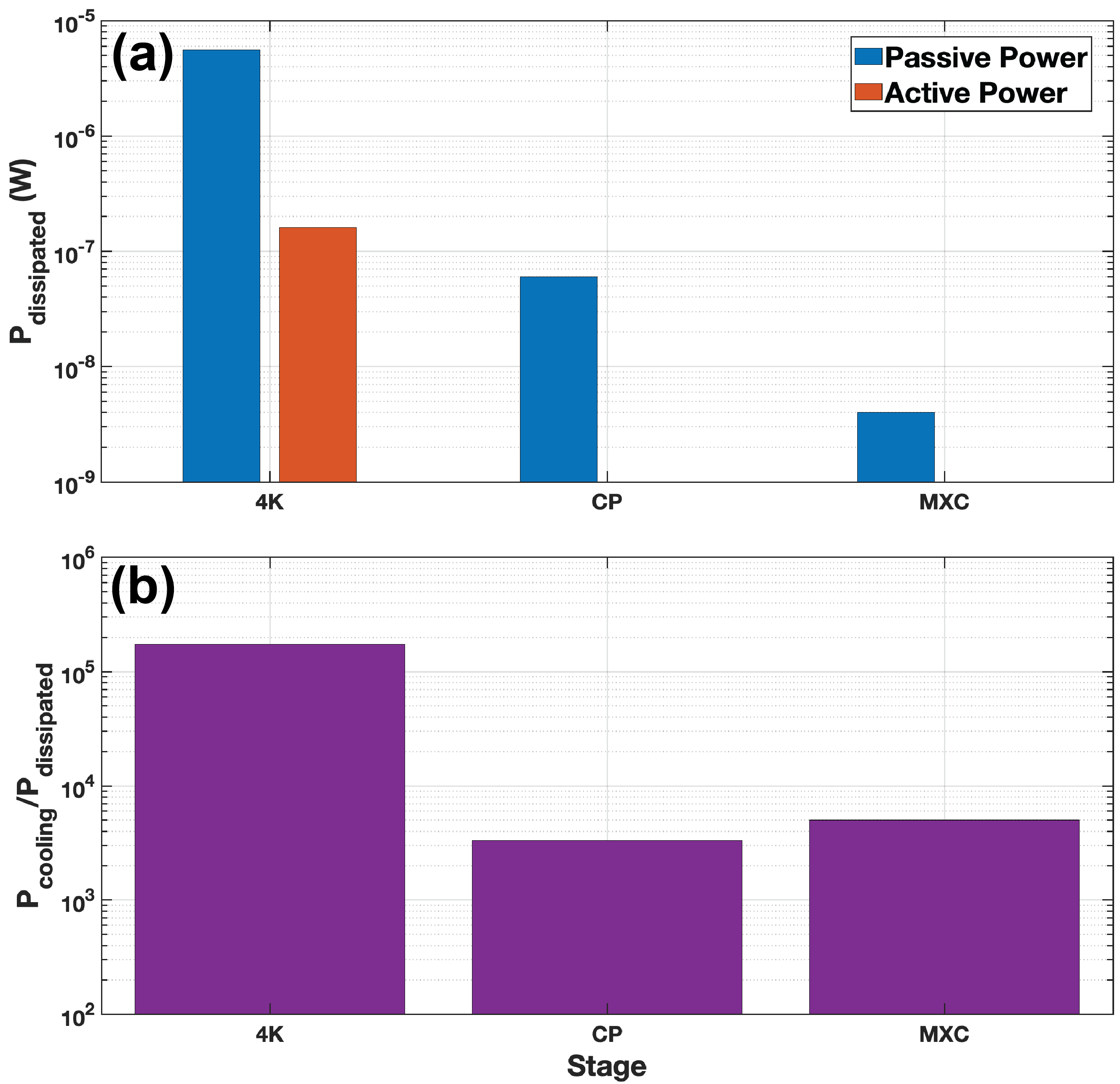}
        \caption
        {
           a) Active and passive power dissipated (heat load) of each stage, b) The  cooling power divided by the sum of the active and passive dissipated power for each stage.
        }
        \label{fig:barchartpp}
    \end{center}
      \vspace*{-\baselineskip}

\end{figure}

Figure~\ref{fig:barchartpp}a shows the active and passive dissipated power ($P_{dissipated}$) for each stage. Next, we have calculated the ratio of cooling power to the total heat load ($P_{cooling}/P_{dissipated}$) as plotted in Fig.~\ref{fig:barchartpp}b
using Table~\ref{tab:coolingpower}. This is an important metric because it can be used to determine the maximum number of cables/fibers that can be supported without exceeding the heat budgets. We can assume that these numbers set the upper limit on the number of qubits that can be operated in a dilution fridge. We also note that the actual heat load contribution in each stage depends on the the duty-cycle of control signals during the runtime of a quantum processor~\cite{Lecocq2021}. Although control signal duty-cycles of less than 10\% have been used in today's demonstrations~\cite{Arute2019,Andersen-natureph2020}, this number will be larger for practical QCs with faster gate speeds. As a result, we have decided to assume an activity rate of 33\% similar to~\cite{Krinner2019,Barends2014} in the next section. 


\section{Discussion and Future Directions}
\label{sec:discussion}

The results from Fig.~\ref{fig:barchartpp}b show that the passive heat load of the CP stage will limit the total number of XY control lines to $\sim$3,000 in our proposed approach. Since the heat budget of 4K stage allows for having more than $10^5$ control signals, further reduction of the passive heat loads of CP and MXC stages by using more advanced SC wiring can increase the capacity. For instance, advanced SC wires ~\cite{smith2021,Tuckerman2016,ACTPol2016} have shown promising paths toward lowering passive heat in cryogenic stages by at least a factor of three. Additionally, our approach will be compatible with potential frequency-multiplexing schemes~\cite{patra-jssc2017} to control multiple qubits over a single wavelength RF-photonic link. This can provide an additional scalability factor in the future.

\begin{figure}[b]
    \begin{center}
        \includegraphics[width=0.45\textwidth, keepaspectratio]{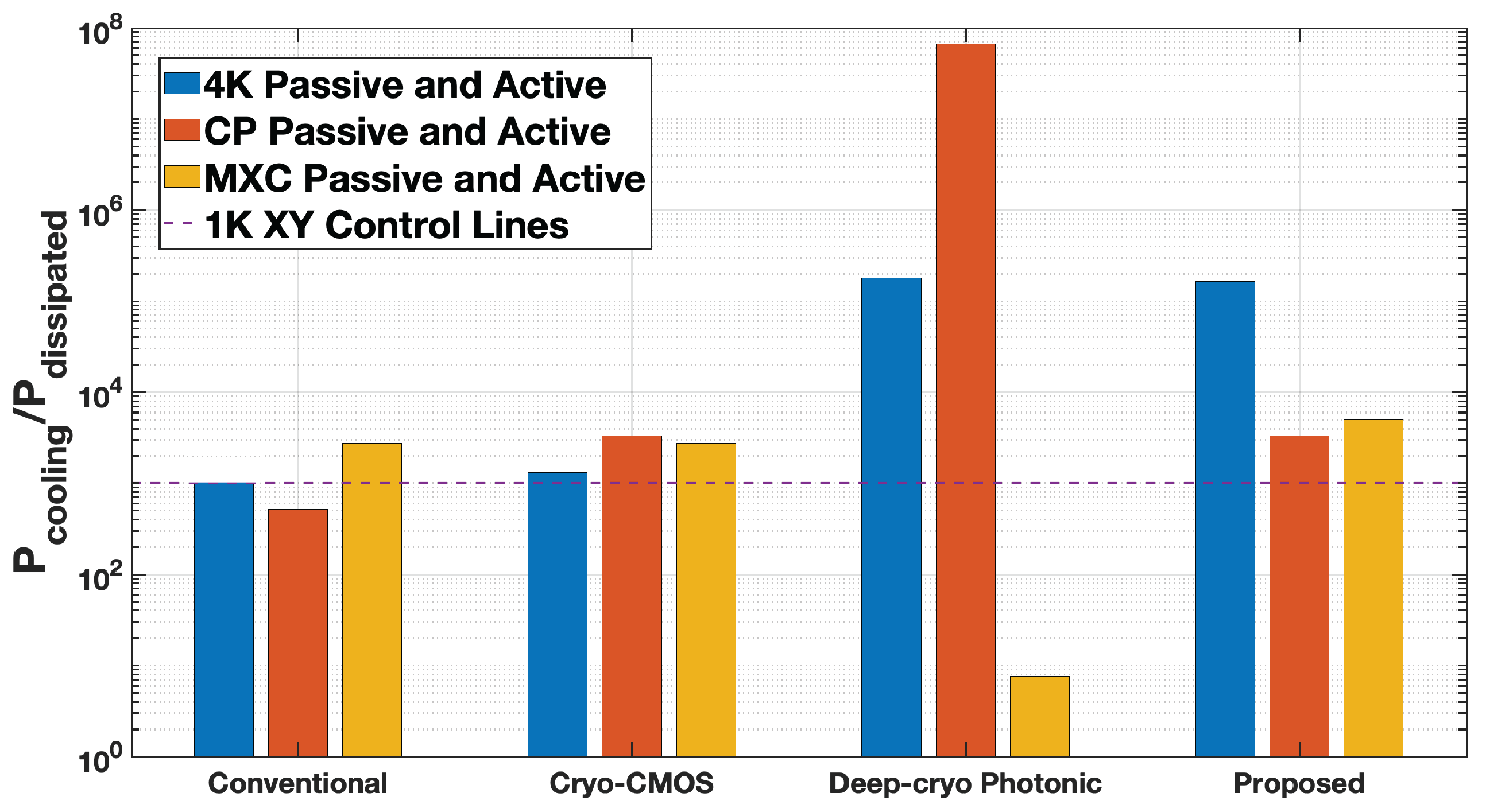}
        \caption
        {
           Comparison plot of maximum number of XY control lines allowed for each stage in the conventional, cryogenic CMOS, deep-cryogenic photonic and our proposed approach.
        }
        \label{fig:comparison}
    \end{center}
      \vspace*{-\baselineskip}

\end{figure}

Figure~\ref{fig:comparison} compares the performance limits in terms of ($P_{cooling}/P_{dissipated}$) among all the state-of-the-art methods (from Section~\ref{sec:implementations}) and our proposed method assuming the duty cycle of 33\%. While we are only considering the XY lines in this comparison (ignoring the Z-flux lines and readout signals), the plot can provide an upper-bound limit on the total number of qubits that can be controlled. This limit can be determined by the lowest $P_{cooling}/P_{dissipated}$ ratio among three stages for each method. 
The number of qubits in a conventional method will ultimately be limited to $\sim1000$ due to the passive heat load of RF coax cables in the 4K stage. Cryogenic CMOS demonstrations have been able to slightly exceed this limit by reducing the passive power in cost of $>2mW$/qubit active power overhead due to cryogenic circuitry. While this power overhead can be slightly reduced by using more advanced CMOS process nodes~\cite{Hung-esscirc2021,Parvais-essderc2018}, it will be very challenging to reduce it by an order of magnitude, as the analog/RF circuit power starts dominating, and it does not scale well with technology node. The deep-cryogenic method overcame the heat budget limitations of the 4K stage by moving all the active components (i.e., photodiode in this approach) down to the MXC stage. However, the stringent cooling power of this stage ($\sim20 \mu W$) only allows for the control of a few qubits. In addition, we assumed $3pW$ passive heat load for CP and MXC stages as claimed in this work while it has not been experimentally verified.

\begin{figure}[b]
    \begin{center}
        \includegraphics[width=0.45\textwidth, keepaspectratio]{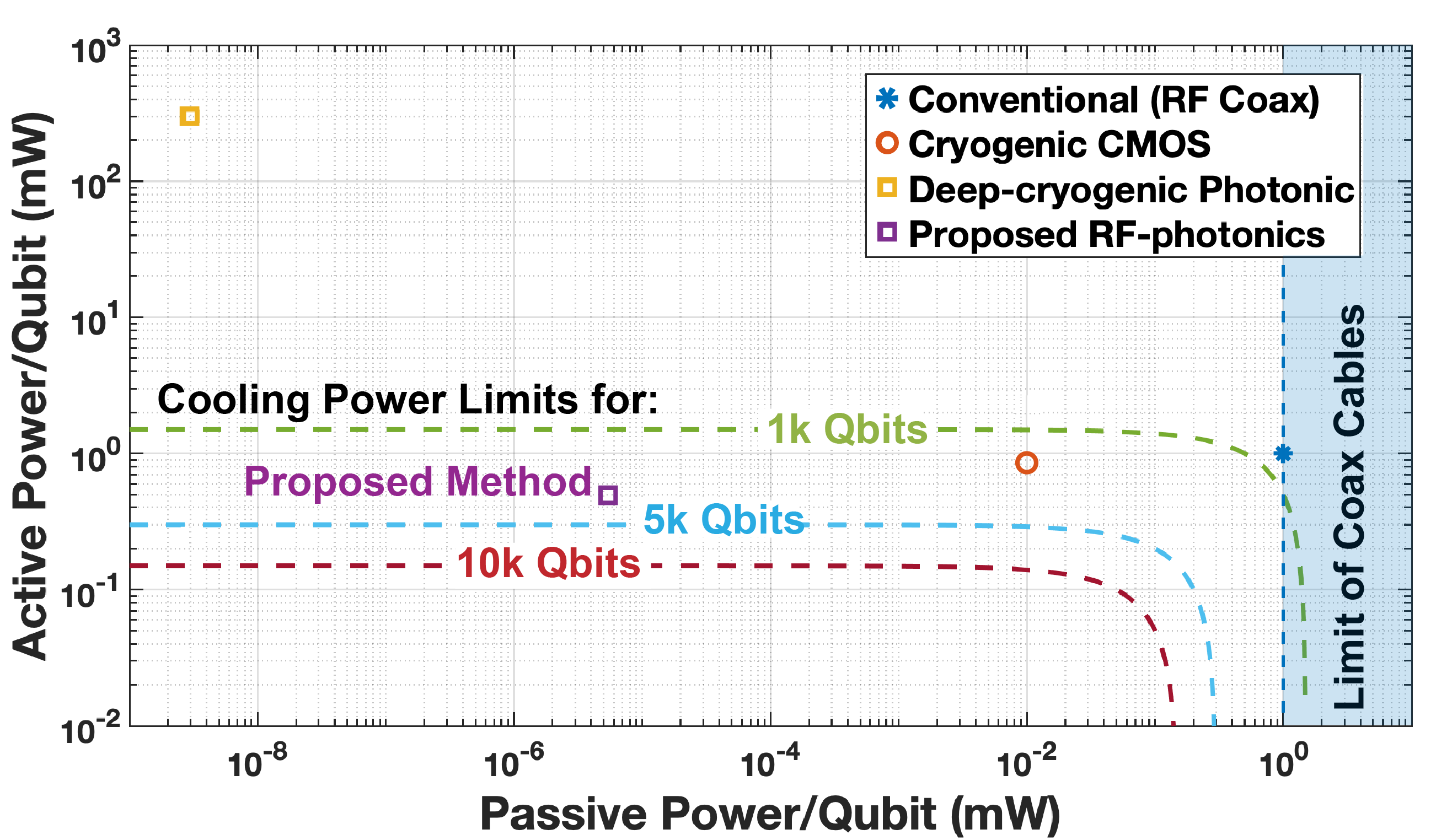}
        \caption
        {
           Active and passive power per qubit for different approaches (assuming a single fiber/wire is required to the control a qubit).
        }
        \label{fig:actpaspower}
    \end{center}
      \vspace*{-\baselineskip}

\end{figure}

As mentioned before, our proposed method can achieve the largest $P_{cooling}/P_{dissipated}$, over $3000$, by exploiting RF-photonic links, while balancing the heat load across the cryogenic stages. This is accomplished by performing the electro-optical conversions at the 4K stage and relying on SC interconnects to interface with qubits. We can summarize this discussion by plotting the active and passive power per qubit as plotted in Fig.~\ref{fig:actpaspower}. In this figure, we have scaled the limiting heat load of each method to the 4K stage by normalizing the cooling power of that stage to the 4K stage.

\begin{figure}[t]
    \begin{center}
        \includegraphics[width=0.45\textwidth, keepaspectratio]{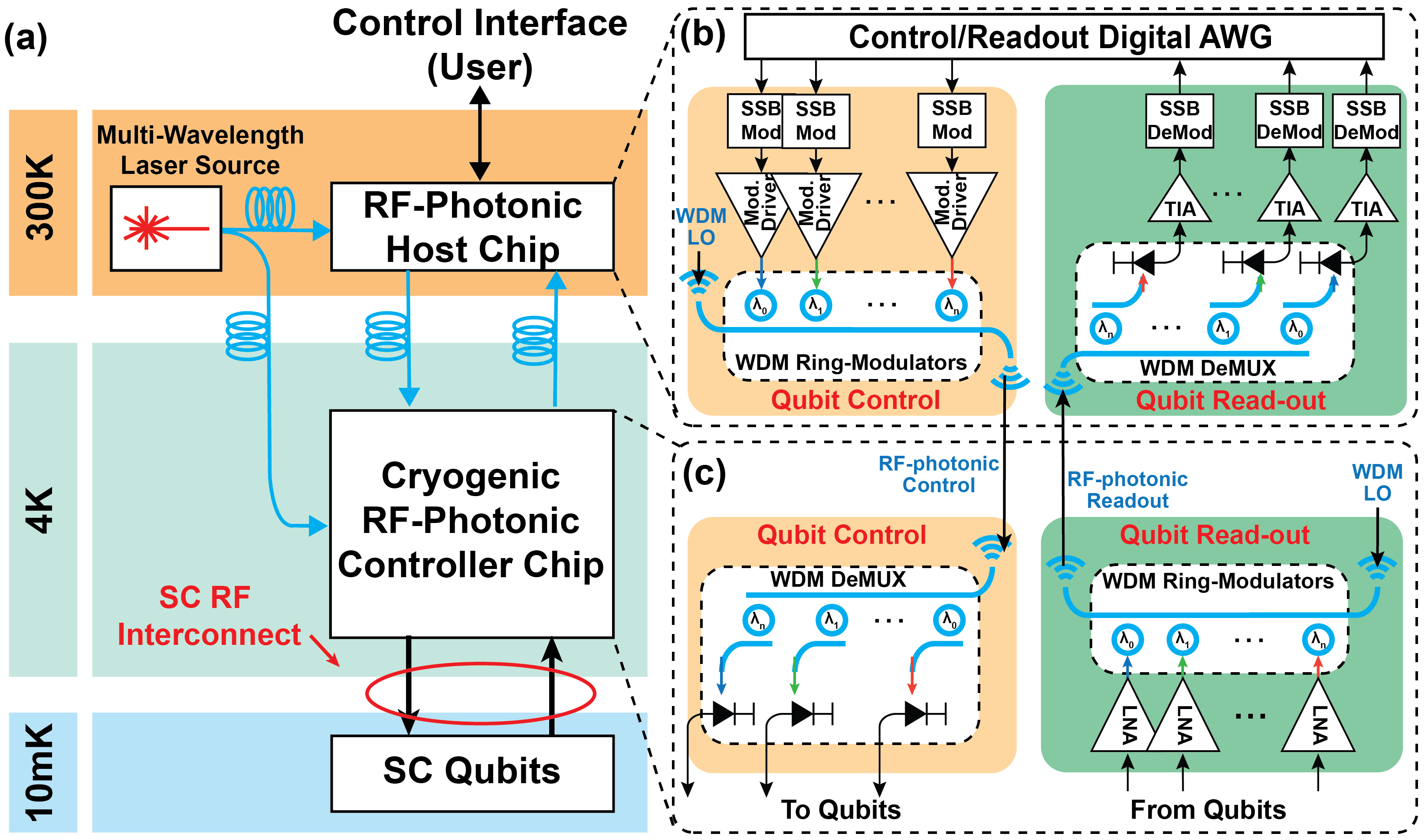}
        \caption
        {
           Proposed realization of a scalable transmon qubit control/readout system using cryogenic RF-photonic in silicon photonics.
        }
        \label{fig:futureidea}
    \end{center}
      \vspace*{-\baselineskip}

\end{figure}

From the implementations perspective, we note that the proposed approach can be efficiently realized using advanced silicon photonics processes. These processes could achieve similar performances at RT and in cryogenic temperatures down to below $\sim4K$~\cite{Gehl-optica2017,Yin-esscirc2021}. Fig.~\ref{fig:futureidea} shows our vision of a scalable RF-photonic approach to control a large number of qubits. This novel architecture comprises of two major components: an RF-photonic host chip at RT (Fig.~\ref{fig:futureidea}b), and a cryogenic RF-photonic controller chip at the 4K stage (Fig.~\ref{fig:futureidea}c). While the host chip can generate precise RF control signals using WDM-compatible optical SSB modulators and receivers, the cryogenic controller chip will contain minimal electro-optical conversion functionality at ultra-low powers. One example implementation of these chips based on ultra energy-efficient ring-modulators~\cite{moazeni-jssc2017,mehta-vlsi2019} is illustrated in Fig.~\ref{fig:futureidea}. However, discussions around detailed design choices and trade-offs are beyond the scope of this paper. 

While the focus of this paper is on the scalable XY control of transmon qubits, we note that Z-flux and coupler control lines, and the readout RF lines (See section~\ref{sec:background}) has to be considered and scaled up in future practical QC system. Extending our proposed scheme for Z-flux lines can be challenging due to the large photocurrent requirements that results in generating excess shot noise at the 4K stage, well above the thermal noise. However, RF-photonic techniques have also shown a promising path for qubit readout: either through electro-optical conversion at the 4K stage~\cite{Youssefi2021}, or through direct microwave-to-optical conversion via intermediary nano-mechanical resonators at the qubit~\cite{Mirhosseini-nature2020}. 
\section{Conclusion}
\label{sec:conclusion}
In this work, we proposed the a new scalable approach that can support XY control of thousands of transmon qubits using cryogenic RF-photonic interconnects. We have studied comprehensive modeling and analysis of noise and thermal budget limitations. By co-optimizing and balancing the passive and active heat loads, presented approach provides XY control of more than $3,000$ qubits for the noisy intermediate-scale quantum (NISQ) era~\cite{Brooks-nature2019}. This method can be realized using current silicon photonics technologies. Finally, the presented RF-photonic approach can be utilized to interconnect various quantum computing platforms~\cite{PRXQuantum.2.017002} and it can be used in extreme-condition physics experiments~\cite{cern-spie2014} as well.

\appendices


\section*{Acknowledgment}
The authors would like to thank Accelerating Quantum-Enabled Technologies (AQET) program at University of Washington for their graduate student fellowship support.

\ifCLASSOPTIONcaptionsoff
  \newpage
\fi



\bibliographystyle{IEEEtran}

\bibliography{emit-refs}



%
\begin{IEEEbiography}[{\includegraphics[width=1in,height=1.25in,clip,keepaspectratio]{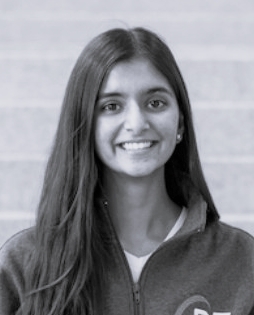}}]{Sanskriti Joshi}
received her B.S. and M.Eng. degree in electrical and computer engineering from Cornell University in 2019 and 2020, respectively. She is currently a Ph.D. student at University of Washington in Seattle, WA. She is the recipient of the 2021 QuantumX AQET fellowship. Her research interests include using integrated circuits and photonics circuits to design scalable quantum computers.
\end{IEEEbiography}

\begin{IEEEbiography}[{\includegraphics[width=1in,height=1.25in,clip,keepaspectratio]{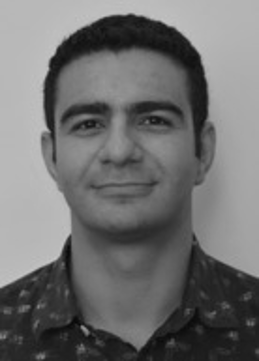}}]{Sajjad Moazeni} (M’13) received the B.S. degree in electrical engineering from the Sharif University of Technology, Tehran, Iran, in 2013, and the M.S. and Ph.D. degrees in electrical engineering and computer science from the University of California at Berkeley, Berkeley, CA, USA, in 2016 and 2018, respectively.

From 2018 to 2020, he was a Post-Doctoral Research Scientist in Bioelectronic Systems Lab at Columbia University, New York, NY, USA. He is currently an Assistant Professor of Electrical and Computer Engineering Department, at University of Washington, Seattle, WA, USA. He received the 2022 NSF CAREER Award. His research interests are designing integrated systems using emerging technologies, integrated photonics, neuro and bio photonics, and analog/mixed-signal integrated circuits.
\end{IEEEbiography}






\end{document}